\documentclass[sigconf]{acmart}

\AtBeginDocument{%
  }
    
\usepackage{svg}

\copyrightyear{2026}
\acmYear{2026}
\setcopyright{cc}
\setcctype{by}
\acmConference[CHI EA '26]{Extended Abstracts of the 2026 CHI Conference on Human Factors in Computing Systems}{April 13--17, 2026}{Barcelona, Spain}
\acmBooktitle{Extended Abstracts of the 2026 CHI Conference on Human Factors in Computing Systems (CHI EA '26), April 13--17, 2026, Barcelona, Spain}
\acmDOI{10.1145/3772363.3798700}
\acmISBN{979-8-4007-2281-3/2026/04}

\begin{document}

\title{CASCADE A Cascading Architecture for Social Coordination with Controllable Emergence at Low Cost}

\author{Yizhi Xu}
\email{2025050012@mails.szu.edu.cn}
\orcid{0009-0001-8958-0122} 
\affiliation{%
  \institution{Shenzhen University}
  \city{Shenzhen}
  \country{China}
}

\renewcommand{\shortauthors}{Xu}

\begin{abstract}
Creating scalable and believable game societies requires balancing authorial control with computational cost. Existing scripted NPC systems scale efficiently but are often rigid, whereas fully LLM-driven agents can produce richer social behavior at a much higher runtime cost. We present \textbf{CASCADE}, a three-layer architecture for low-cost, controllable social coordination in sandbox-style game worlds. A \textbf{Macro State Director} (Level 1) maintains discrete-time world-state variables and macro-level causal updates, while a modular \textbf{Coordination Hub} decomposes state changes through domain-specific components (e.g., professional and social coordination) and routes the resulting directives to tag-defined groups. Then \textbf{Tag-Driven NPCs} (Level 3) execute responses through behavior trees and local state/utility functions, invoking large language models only for on-demand player-facing interactions. We evaluate CASCADE through multiple micro-scenario prototypes and trace-based analysis, showing how a macro event can produce differentiated yet logically constrained NPC behaviors without per-agent prompting in the main simulation loop. CASCADE provides a modular foundation for scalable social simulation and future open-world authoring tools.
\end{abstract}




\maketitle

\section{Introduction}

Recent attempts to build “living” game societies increasingly rely on LLM-driven NPCs that generate dialogue and action in an agent-centric manner. Although this approach can improve local behavioral expressiveness, it introduces three practical problems for game-scale deployment: high runtime cost and technical barriers, difficulty maintaining coherent world logic under stochastic generation, and increased exposure to player-driven prompt exploitation or other unsafe outputs. In practice, these systems often make it difficult to sustain large populations while preserving authorial control over the broader simulation.

This paper argues that the central design problem is not how to make every NPC more intelligent, but how to organize social behavior at a level that remains scalable, steerable, and interpretable. Instead of centering computation around per-agent generation, we explore a different design unit: social coordination. Our goal is therefore not to maximize individual realism at all costs, but to provide a lower-cost and more controllable architecture for orchestrating population-level behavior in game worlds.

Consider a drought event in a small town. In CASCADE, a modular Coordination Hub decomposes the macro condition through domain-specific components (e.g., profession, social coordination, governance, or resource management), each of which produces different classes of group directives, such as water-conservation signals for farmers and price-adjustment signals for merchants. At the local level, NPCs do not respond identically: one merchant may raise prices, while another distributes aid, depending on tags, needs, and local state. This end-to-end flow illustrates the central idea of CASCADE: transforming a shared macro condition into differentiated yet constrained social behavior without continuous per-agent prompting.

 \textbf{CASCADE} shifts control from agent centric generation to coordination centric simulation. It separates macro-level causal updates, meso-level modular coordination, and micro-level local execution, enabling a design trade-off that favors controllability, modularity, and lower runtime cost over fully generative per-agent. Our contributions are threefold: (1) a coordination-centric alternative to mainstream LLM-driven NPC architectures, (2) a layered mechanism for low-cost social simulation through cascading directives and local policies, and (3) a modular prototype framework that demonstrates how domain-specific coordination modules can support extensible and designer-controllable social simulation.

\newpage

\section{Related Work}

We situate CASCADE in the context of game AI pipelines, simulated social agents, and scalable coordination mechanisms. Below we review existing approaches and their limitations.Research on game AI and social simulation spans a wide spectrum, from traditional scripted control to recent attempts at leveraging large language models. In conventional game development, techniques such as behavior trees, finite state machines, and utility-based systems have been widely adopted for NPC control due to their efficiency and predictability. These approaches enable designers to implement concise and debuggable behaviors, but they typically require substantial manual authoring and do not generalize well to complex social dynamics. Integrating advanced generative mechanisms into these pipelines remains challenging, as the structured nature of traditional systems and the unstructured outputs of generative models are difficult to reconcile in real-time production environments.

Recent work has explored the use of large language models to drive individual agent behavior in simulations. Such LLM-driven agents can produce more expressive dialogue and context-aware actions, and have been demonstrated in settings ranging from small-group interaction studies to research prototypes. One prominent example is \textit{Generative Agents} \cite{Park2023Generative}. However, in game-scale environments these systems face practical limitations: per-agent prompting can be computationally expensive, generated outputs may drift from intended world logic, and the lack of explicit control can make it difficult to enforce consistent narrative constraints. Prior research in generative agents often focuses on individual cognitive modeling rather than scalable population-level coordination.
Industrial game engines and commercial titles have experimented with hybrid AI pipelines that combine lightweight scriptable decision frameworks with selective calls to generative components\cite{Vezhnevets2023Concordia}. These solutions can incorporate emergent-like behavior in limited contexts while retaining performance, but they tend to be deeply integrated with proprietary tooling and tailored to specific genres or gameplay mechanics. Standardizing such patterns for broader reuse remains an open challenge.

In contrast to agent-centric generative designs and tightly coupled industrial pipelines, CASCADE proposes a coordination-centric architecture that separates macro-level state management from meso-level directive routing and micro-level local execution. By shifting the locus of complexity from per-agent generation to structured coordination layers, CASCADE enables a different trade-off between expressiveness, control, and runtime cost. This distinction sets it apart from existing approaches while aligning with the practical needs of scalable social simulation in games.

\section{The CASCADE Architecture}

\subsection{Overview}
CASCADE is a three-layer architecture designed to enable controllable, social emergence at a low computational cost. It separates macro-level causal updates, meso-level directive routing, and micro-level local execution, enabling a design trade-off that favors modularity, low runtime cost, and scalability while maintaining coherent narrative control. This section provides a detailed breakdown of each layer's components, functionalities, and the workflows within the system.

\begin{figure}
    \centering
    \includegraphics[width=0.75\linewidth]{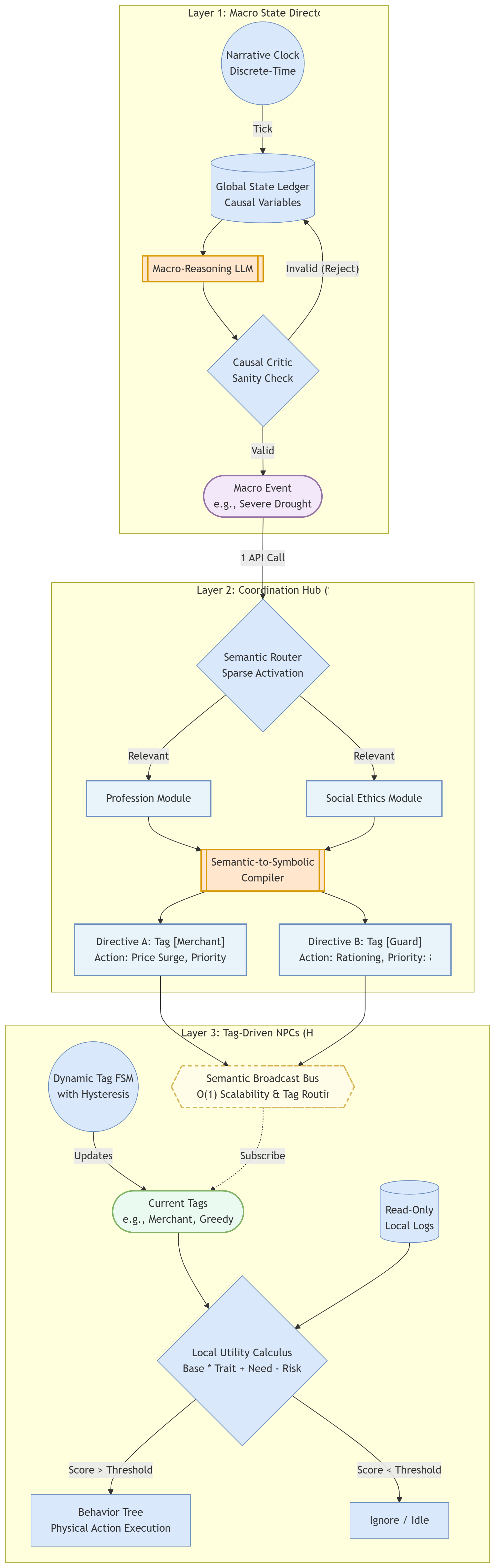}
    \caption{CASCADE Architecture: A Macro State Director drives the simulation, filtered through a Coordination Hub, and executed by Tag-Driven NPCs via local utility functions.}
    \label{fig:architecture}
\end{figure}

\subsection{Layer 1: Macro State Director}
The Macro State Director operates at the macro level, maintaining a discrete-time causal system for the game world. It is responsible for triggering significant global events, which are propagated to the Coordination Hub for further processing. The key components of Layer 1 include:

Narrative Clock: This is the heart of the Macro State Director, driving discrete-time steps (e.g., each in-game day) and ensuring that world states are updated only at specific time intervals. This guarantees that computational load remains manageable, as only significant state changes trigger event processing.

Global State Ledger: The Ledger stores the "objective truth" of the game world, tracking causal variables like "Water Scarcity" or "Political Stability." It does not directly track NPC behaviors but provides a record of high-level state transitions over time. This history is crucial to understanding the progression of the world.

Macro-Reasoning Engine: The engine processes global state updates, applying rules or LLMs to evaluate potential macro-events. If certain thresholds are reached (e.g., "Water Scarcity" reaches a critical level), the engine triggers a new event (e.g., "Drought").

Causal Critic: This component ensures that all macro-events align with the historical and causal constraints of the game world. If a proposed event (such as a drought) conflicts with the current state (e.g., the world is in a rainy season), the Causal Critic flags it as inconsistent and prevents it from being processed.

\subsection{Layer 2: Coordination Hub}
Layer 2 serves as the modular coordination center, responsible for translating macro-level state changes into concrete group-level directives. The Hub interacts with various domain-specific modules, each focusing on a different aspect of the game world, and generates a structured output (in the form of JSON packets). These directives are routed to NPCs using a tag-based routing system.

Sparse Activation: The Coordination Hub's semantic router evaluates each macro-event and activates only the relevant modules. For example, if "Water Scarcity" reaches a critical level, only the "Resource Allocation" and "Security" modules may be activated, leaving other modules (such as entertainment) inactive.

Compilation and Translation: Once a module is activated, it converts the abstract state change (e.g., "Water Scarcity: Critical") into specific, actionable directives. These directives include parameters that are domain-specific (e.g., increasing water prices for merchants or organizing water patrols for guards).

Tag Routing: The directives are broadcast to relevant NPCs by tags (e.g., all NPCs with the [Merchant] or [Farmer] tag). This ensures that only the relevant NPCs receive the right instructions, minimizing computational overhead while maintaining high modularity.

\subsection{ Layer 3: Tag-Driven NPCs}
The third layer is where NPCs execute their behaviors based on directives from the Coordination Hub. NPCs are defined by tags (e.g., [Merchant], [Farmer], [Guard]), which determine their roles and the channels they subscribe to. Each NPC utilizes a local utility function to assess the directive and decide on the appropriate behavior.

Dynamic Tag Migration: Tags are not static. An NPC may change tags (e.g., from [Merchant] to [Beggar]) depending on local variables, such as wealth. This dynamic tag migration is handled by a lightweight finite state machine (FSM), ensuring that NPCs can adapt to changes in the game world, such as a sudden loss of resources.

Local Utility Calculus: When an NPC receives a directive, it evaluates whether the directive aligns with its needs, traits, and goals. The utility function incorporates factors such as base priority, the NPC’s traits (e.g., greed), current needs (e.g., hunger), and risks (e.g., danger from other NPCs). If the utility score exceeds a threshold, the NPC will perform the associated action.

Action-Dialogue Decoupling: The NPC’s behavior and dialogue are kept separate. While NPC actions (e.g., “patrol,” “hoard”) are determined by local utility functions, dialogue is only generated when needed (e.g., during player interactions) to prevent excessive computational cost from constant dialogue generation.

\subsection{Summary}
By decoupling macro-level causal events from individual NPC actions, the architecture balances computational efficiency with narrative control, allowing game designers to manage large populations of NPCs with minimal overhead, while preserving the richness and diversity of social behaviors. This design offers a flexible foundation for creating complex, dynamic game worlds that can respond to a variety of global events and interactions, opening new possibilities for scalable, controllable social simulation.

\section{Evaluation}

To evaluate the performance and behavior of CASCADE, we conducted several micro-scenario prototypes, demonstrating how macro level events (such as a drought) lead to diverse but logically constrained NPC behaviors. These experiments validate the scalability and controllability of CASCADE by testing different settings and configurations.

\subsection{Experimental Setup}

The evaluation was conducted using a simulated town consisting of 10 NPCs with various roles (e.g., farmers, merchants, and a mayor). We used a modular setup to demonstrate how a macro event, such as a severe drought, can trigger differentiated behaviors based on each NPC's tag and local conditions. The experimental setup includes:

\begin{itemize}
    \item A \textbf{Macro event}: Severe Drought, which affects the water scarcity state.
    \item A \textbf{Coordination Hub}: Translates the macro event into domain-specific directives, such as price adjustments for merchants and rationing for guards.
    \item \textbf{NPC Responses}: NPCs act based on local utility functions, influenced by their tags, needs, and state.
\end{itemize}

\begin{table*}[t]  
    \caption{NPC Actions During the "Severe Drought" Event}
    \label{tab:npc_actions}
    \centering
    \begin{tabular}{l l l}
    \toprule
    \textbf{NPC} & \textbf{Tags} & \textbf{Action} \\
    \midrule
    Mayor & [Leader], [Lawful] & Suggest water conservation, initiate town hall meeting \\
    Merchant 1 & [Merchant], [Greedy] & Increase water price by 30\% \\
    Merchant 2 & [Merchant], [Generous] & Offer discounted water to the poor \\
    Farmer 1 & [Farmer], [Hardworking] & Collect and ration water for crops \\
    Farmer 2 & [Farmer], [Lazy] & Ignore water distribution orders, idle \\
    Guard 1 & [Guard], [Responsible] & Patrol water sources and enforce rationing \\
    Guard 2 & [Guard], [Lazy] & Skip patrols, idle \\
    \bottomrule
    \end{tabular}
\end{table*}

\newpage

\subsection{Results: Macro Event Impact on NPC Behavior}

The evaluation demonstrated that CASCADE can effectively simulate the social dynamics of a small town. The macro event of "Severe Drought" caused differentiated responses among NPCs without per-agent prompting. The following table summarizes the actions taken by NPCs in response to the drought event:

\subsection{Behavior Diversity and Cost Comparison}

The table demonstrate how the CASCADE architecture successfully coordinates diverse behaviors, providing both the scalability and the controllability necessary for large-scale simulations. 

We compared the LLM usage between the CASCADE model and a baseline "Full-Generative" NPC model that relies on constant per-agent prompting. In CASCADE, NPCs only invoke LLMs for player-facing dialogue, leading to a drastic reduction in token cost.

\subsection{Scalability and Performance Evaluation}

To evaluate the scalability of CASCADE, we conducted simulations with varying numbers of NPCs and tracked both performance and behavioral diversity. As the number of NPCs increases, CASCADE's performance remains steady, and the system continues to produce diverse behaviors without a significant increase in computational cost.

\subsection{Conclusion of Evaluation}

The evaluation results demonstrate that CASCADE offers a scalable and controllable framework for social coordination in game worlds. By decoupling macro-level causality, meso-level directive routing, and micro-level execution, CASCADE effectively maintains control over emergent behaviors while minimizing computational cost. The experiments confirmed the viability of using CASCADE in large-scale simulations without the need for expensive per-agent prompting.

\section{Discussion}The CASCADE architecture demonstrates a viable pathway out of the "Simulation Paradox" that currently plagues generative game environments. By shifting the paradigm from agent-centric reasoning to coordination-centric orchestration, we open new possibilities for both game designers and HCI researchers.
\subsection{Design Implications: Towards a "Social Physics" Engine}Current generative agent frameworks operate under a "One-Agent-One-Model" assumption, forcing individual NPCs to shoulder the entire cognitive burden of social coordination. This not only results in prohibitive $O(N)$ computational scaling but also strips game directors of narrative control. CASCADE resolves this by functioning as a "social physics" engine. Just as a traditional physics engine handles gravity and collisions universally without requiring each object to compute the laws of thermodynamics, CASCADE's \textbf{Coordination Hub} handles semantic-to-symbolic compilation at a societal level. This allows designers to easily author macro-level fate while trusting the micro-level execution to handle localized, zero-cost diversity.Furthermore, CASCADE introduces a critical security paradigm: \textbf{Action-Dialogue Decoupling}. In pure LLM-driven systems, players can use prompt injection to trick agents into executing destructive or out-of-character actions. Because CASCADE confines the LLM to a read-only conversational interface grounded in symbolic tags, player manipulation cannot bypass the local utility calculus to overwrite physical behavior. This neuro-symbolic isolation provides the safety guarantees necessary for deploying open-ended AI in commercial environments.

\subsection{Limitations and Future Work}While the current implementation successfully orchestrates top-down cascading directives, it exhibits limitations that pave the way for a more organic "CASCADE 2.0" ecosystem:\textbf{Bidirectional Causality and Bottom-Up Emergence:} The current architecture relies on a unidirectional flow (L1 $\rightarrow$ L2 $\rightarrow$ L3). If a subset of NPCs forms a riot due to local utility conflicts, this micro-level anomaly does not organically rewrite the L1 world state. Future iterations will introduce a \textbf{Micro-Anomaly Aggregator} utilizing Complex Event Processing (CEP). By calculating statistical thresholds of low-level symbolic events (e.g., high density of "Brawl" actions), the system will automatically synthesize macro-events and push them upward, achieving true bidirectional societal evolution without linear token costs.\textbf{Social Topology and Spatial Damping:} Currently, the Semantic Broadcast Bus routes directives uniformly to all subscribed tags. Real societies, however, rely on topological networks where information decays over distance and trust . Future work will replace pure broadcasting with Top-K semantic routing and integrate SIR (Susceptible-Infectious-Recovered) epidemiological models to simulate the spatial-temporal attenuation of rumors and directives.

\textbf{Towards an Open-Source Ecosystem and Extensible Authoring:} Ultimately, our vision extends beyond providing a single architectural framework. We aim to evolve CASCADE into an open-source, community-driven platform. While initial authoring tools will focus on coordination and performance, the long-term goal is to build an extensible ecosystem akin to a modular plugin repository. Game developers and researchers will be able to construct and share their own Macro State Directors (Layer 1) tailored to diverse game genres. Furthermore, the Coordination Hub (Layer 2) will support an open library of pluggable, domain-specific modules---expanding beyond basic economics and security to encompass complex societal facets like entertainment, political governance, kinship, and religion. By crowdsourcing these meso-level modules, CASCADE aims to serve as a standardized, highly realistic ``social physics'' middleware for the next generation of interactive digital societies.

\bibliographystyle{ACM-Reference-Format}
\bibliography{myrefs}

\end{document}